\documentstyle[prl,twocolumn,aps,floats]{revtex}
\input{epsf}

\begin{document}
\draft
\twocolumn
\title{Photocount statistics of chaotic lasers}

\author{G.\ Hackenbroich, C.\ Viviescas, B.\ Elattari, and F.\
  Haake}
\address{Universit\"at Essen, Fachbereich 7, 45117 Essen,
  Germany}

\maketitle

\begin{abstract}
  We derive the photocount statistics of the radiation emitted from
  a chaotic laser resonator in the regime of single--mode lasing.
  Random spatial variations of the resonator eigenfunctions lead to
  strong mode--to--mode fluctuations of the laser emission. The
  distribution of the mean photocount over {\em an ensemble of
    modes} changes qualitatively at the lasing transition, and
  displays up to three peaks above the lasing threshold.
\end{abstract}
\vspace*{-0.05 truein} 

Nonlinear interactions in active optical media can drastically
affect the statistics of the photon field. The textbook example is
the single--mode laser \cite{Wal94,Man95}: Below the laser
threshold, losses outweigh the gain by linear amplification such
that nonlinear saturation is negligible. Then the photon statistics
is well represented by a thermal distribution, and the probability
$P_n$ for the occupation of the mode with $n$ photons decays as the
power law $P_n \sim [\bar{n}/ (1+\bar{n})]^n$ ($\bar{n}$ is the mean
photon number). Above threshold nonlinear interactions stabilize the
field intensity. The (relative) intensity fluctuations are strongly
reduced below the value found for a thermal distribution, and far
above threshold $P_n$ approaches the Poissonian distribution $P_n =
[\bar{n}^n/n!] \exp(-\bar{n})$ characteristic for a coherent state.
These facts have been known since the development of the quantum
theory of lasing in the 1960s.

Interest in the photon statistics of amplifying media was recently
renewed by experiments on artificially fabricated random media.  In
random media radiation is scattered in an irregular, chaotic way.
Practical realizations include laser dye solutions \cite{Law94} and
semiconductors \cite{Cao99} with randomly fluctuating dielectric
constant as well as chaotic resonators with irregularly shaped
boundaries \cite{Gma98}.  Even in the absence of pumping random
media show interesting interference effects such as coherent
backscattering or localization of light \cite{She90}. The
theoretical investigation of the quantum optical properties of
random media started \cite{Gru96,Art97} with the demonstration of
deviations from blackbody radiation in the case of one dimensional
scattering.  In a pioneering work \cite{Bee98} Beenakker generalized
these results to multidimensional chaotic scattering by establishing
a general relationship between the photocount statistics of a {\em
  linear} random medium and its scattering matrix. Using that
relationship, Beenakker and coworkers predicted an excess photon
noise due to multiple scattering \cite{Bee98}, computed the photon
noise power spectrum \cite{Mis00}, and investigated the effect of
photon localization on the photocount statistics \cite{Bee00}.

In contrast to the wealth of results for linear random amplifiers,
little is known about the photon statistics of random media above
the lasing threshold. This is an important open problem since the
experimental signatures of random lasing are still controversial
\cite{Cao99}. In the present paper we address that problem for a
chaotic laser resonator in the regime of single--mode lasing.  We
show that the chaotic nature of the cavity modes gives rise to
fluctuations of the photocount {\em on top} of the quantum optical
fluctuations known from laser theory. Chaos--induced fluctuations
are found when a single--mode photodetection is performed over an
{\em ensemble} of modes. The ensemble may be obtained from a single
resonator upon varying suitable parameters or from different
resonators with small variations in shape. The factorial moments of
the photocount display strong fluctuations from one mode to another.
However, we show that the probability density for each factorial
moment depends only on general symmetries of the ensemble and on
four parameters describing the laser dynamics; these are the
coefficients of linear gain, mean escape loss, absorption loss and
saturation of the amplifying medium.

\begin{figure}[t]
\begin{center}
\leavevmode
\vspace*{-1.1cm}
\epsfxsize = 7.2cm
\epsfbox{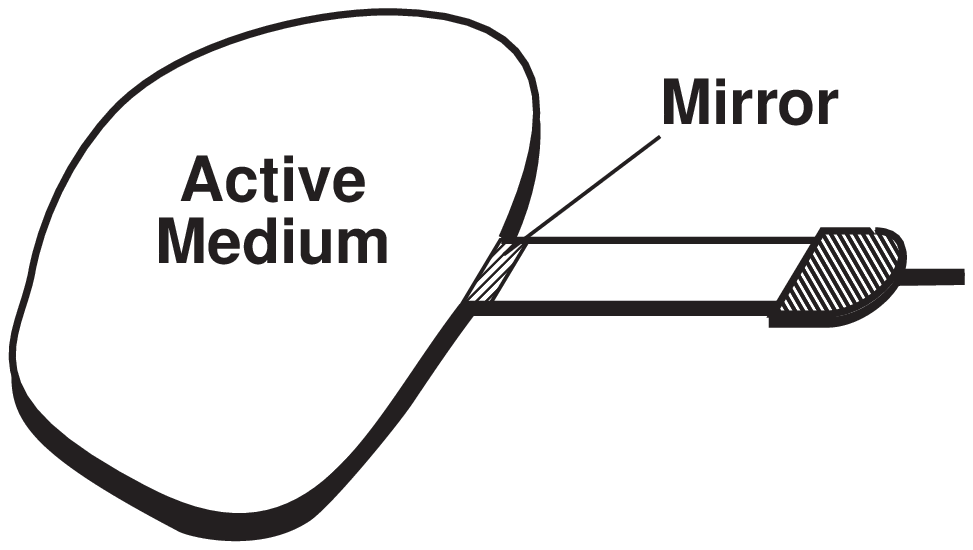}
\end{center}
\vspace*{0.9cm}
\caption{Sketch of chaotic laser cavity with a partly transmitting
  mirror. The cavity is connected to a waveguide and a frequency
  selective photodetector.}
\end{figure}

A sketch of the system is shown in Fig.~1: A chaotic resonator with
an irregularly shaped boundary is homogeneously filled with an
active medium. A partly transmitting mirror reflects impinging
radiation with mean probability $R$ back into the resonator. The
fraction $T=1-R$ of radiation is transmitted and injected into a
waveguide supporting $M$ transverse modes. A frequency selective
photodetector counts the transmitted photons with efficiency $1$. We
consider the case of a quasi discrete cavity spectrum; then the mean
cavity escape rate $\overline{\Gamma} = MT \Delta \omega$ is smaller
than the spacing $\Delta \omega$ of adjacent resonator modes.
Accordingly, the mirror transmission coefficient $T$ must be small,
$MT \ll 1$. The counting time $t$ will be assumed large enough that
the radiation from individual cavity modes can be resolved, $t
\Delta \omega \gg 1$.

As for the active medium, we allow for $N$ active atoms. The $\mu$th
atom ``sees'' the field mode through a coupling constant $g_\mu$
which is proportional to both the atomic dipole matrix element and
the value of the lasing mode at the location of the atom. We only
consider the simplest situation in which the characteristic times
for atomic pump and losses are short compared to the mean life time
of a photon in the cavity. The effect of the atoms on the field mode
can then be represented by three parameters ${\cal A}$, ${\cal B}$,
${\cal C}$ that characterize the linear gain, the nonlinear
saturation, and the total loss, respectively \cite{Man95}. The first
two of these depend on the coupling constants, ${\cal A}\sim
\sum_{\mu=1}^Ng_{\mu}^2$, ${\cal B}\sim\sum_{\mu=1}^Ng_{\mu}^4$.
Inasmuch as the resonator mode represents wave chaos, it varies
quasi-randomly on the scale of the optical wavelength $\lambda$
\cite{Berry,Bibel}. Coupling constants $g_\mu$ associated with atoms
separated by more than $\lambda$ therefore behave as independent
random numbers. The coefficients ${\cal A}$ and ${\cal B}$, however,
become sharp in the limit $N \gg 1$, due to the central limit
theorem. The total loss rate ${\cal C}=\Gamma+\kappa$ is the sum of
the photon escape rate $\Gamma$ and the absorption rate $\kappa$.
Here, the absorption rate may be considered fixed while the escape
rate is a random quantity, as will be revealed presently.

To calculate the photocount statistics we use the input--output
theory of Gardiner and Collet \cite{Gar85,Wal94}. Waves entering and
leaving the waveguide are described by $M$--component annihilation
and creation operators $a^{\rm in}$, $a^{{\rm in}\dagger}$, $a^{\rm
  out}$, $a^{{\rm out}\dagger}$, which obey the commutation
relations
\begin{eqnarray}
[a_p(t),a_q^\dagger(t^\prime)] =
\delta_{pq}\delta(t-t^\prime),\quad \;
[a_p(t),a_q(t^\prime)] &= & 0\,.
\end{eqnarray}
Here, $p,q=1,\ldots,M$ label the transverse modes and $a=a^{\rm
in}$ or $a=a^{\rm out}$. The boundary condition
\begin{eqnarray}
a_p^{\rm in} (t) + a_p^{\rm out}(t)=\gamma_p b(t)\, ,
\label{eq:bound}
\end{eqnarray}
connects incoming and outgoing radiation in each transverse mode
with the annihilation operator $b$ of the cavity mode. While the
boundary condition (\ref{eq:bound}) results from assuming a linear
coupling between the waveguide and the cavity field, {\em no}
restriction is imposed on the intracavity dynamics.  Therefore, the
input-output relation (\ref{eq:bound}) holds both below and above
threshold. The outcoupling amplitudes $\gamma_p$ must be considered
independent random quantities with Gaussian statistics, since they
represent the local fluctuations of the resonator mode across the
outcoupling mirror. In fact, their randomness constitutes the
principle effect of the wave chaos within the resonator on the
ensemble fluctuations of the laser output.

In the linear regime below threshold, one can eliminate the cavity
operators from Eq.\ (\ref{eq:bound}) and express the outgoing
radiation in terms of the incoming radiation, the intracavity noise,
and the S--matrix \cite{Wal94,Bee98}. One thus obtains the
photocount fluctuations \cite{Bee98} through the S--matrix
statistics known for chaotic scattering.

We proceed to the nonlinear regime near and above threshold. For the
case of a vacuum input, the boundary condition (\ref{eq:bound})
allows to express all variances of the output field entirely in
terms of the variances of the cavity field and the Gaussian
statistics of the $\gamma_p$.  We characterize the photocount
statistics through the factorial moments
\begin{eqnarray}
\mu_r = \sum_{m=0}^\infty m(m-1)\cdots(m-r+1)p(m)
\label{eq:fact1}
\end{eqnarray}
of the probability $p(m)$ that $m$ photons are counted in a time
interval $t$. That probability is given by \cite{Wal94,Man95}
\begin{eqnarray}
\nonumber
p(m) & = &{1 \over m!} \langle :W^m e^{-W}: \rangle\,, \\
\label{eq:W} W \  &=& \int_0^t d t^\prime \sum_{p=1}^M
a_p^{{\rm out} \dagger}
(t^\prime) a_p^{\rm out} (t^\prime)\,.
\end{eqnarray}
Here $\langle \cdots \rangle$ denote the quantum steady--state
average and the colons demand normal and a certain time ordering.
Combining Eqs.\ (\ref{eq:fact1})-(\ref{eq:W}) with the input--output
relation (\ref{eq:bound}) for a vacuum input we get
\begin{eqnarray}
\mu_r  =  \langle:W^r \! :\rangle = \Gamma^r \langle: I^r \!
:\rangle\, ,
\label{eq:fact2}
\end{eqnarray}
where $\Gamma = \sum_{p=1}^M |\gamma_p|^2$ is the escape rate and
$I^r$ the $r$th power of the integrated cavity field intensity
\begin{eqnarray}
\label{eq:intens}
I   =   \int_0^t dt^\prime b^\dagger (t^\prime) b(t^\prime)\, .
\end{eqnarray}
Clearly, the first moment alias mean photocount, $\mu_1=\Gamma
t\langle b^\dagger b\rangle$, is a purely static quantity as it is
proportional to the stationary mean photon number inside the cavity.

The factorial moments (\ref{eq:fact2}) specify the output statistics
of a single cavity mode with the random escape rate $\Gamma$. The
distribution of $\Gamma$ over an {\em ensemble of modes} is the
$\chi^2_\nu$ distribution
\begin{eqnarray}
P(\Gamma) =A_\nu \Gamma^{\nu / 2 -1} \exp(-\nu \Gamma / 2
\overline{\Gamma})
\label{eq:distgam}
\end{eqnarray}
well--known from random--matrix theory \cite{Guh98}.  Here, $\nu =
\beta M$ is an integer and $A_\nu$ a normalization constant. The
value of the parameter $\beta$ depends on whether the system is
time--reversal invariant ($\beta=1$) or whether time--reversal
invariance is broken ($\beta=2$) \cite{Guh98,beta}. The special case
$M=\beta=1$ is known as the Porter--Thomas distribution.  Together
with $\Gamma$ the factorial moments $\mu_r$ become random
quantities.  Their distribution is given by
\begin{eqnarray}
\label{eq:main}
{\cal P}(\mu_r) = \int d \Gamma P(\Gamma) \delta(\mu_r -
\Gamma^r \langle: I^r \! :\rangle)\, .
\end{eqnarray}
Note that the right hand side involves a twofold average, the
quantum optical average (represented by the brackets $\langle \cdots
\rangle$) and the ensemble average over the cavity modes
(represented by the integral with the probability distribution
$P(\Gamma)$). We emphasize that the quantum optical average
$\langle: I^r \! :\rangle$ depends on $\Gamma$ through the total
loss rate ${\cal C}$. We now discuss the result (\ref{eq:main}) in
various limiting cases.

We first consider the case $M \gg 1$ of many transverse modes in the
waveguide. The diameter of the waveguide is then much larger than
the wavelength of the cavity mode. A simple saddle point argument
shows that $P(\Gamma)$ for large $M$ approaches a Gaussian
distribution with mean $\overline{\Gamma} \sim M$ and standard
deviation $\Delta \Gamma = \overline{\Gamma}/\sqrt{\beta M/2}$. The
relative fluctuations are small, $\Delta \Gamma / \overline{\Gamma}
\sim 1/\sqrt{M} \to 0$ for $M \to \infty$, and the same is true for
the fluctuations of all factorial moments. For cavities with large
outcoupling mirrors we thus recover the sharp factorial moments one
is used to from non-chaotic resonators.

Second, we investigate the limit of vanishing photocount, $\mu_r \to
0$.  According to Eq.\ (\ref{eq:fact2}), this is the weak--coupling
limit $\Gamma \to 0$ for which the photons in the cavity field can
hardly escape into the waveguide. The laser dynamics becomes
independent of the outcoupling loss as the total cavity loss ${\cal
  C}$ is fully dominated by the absorption loss. All cavity field
moments become independent of $\Gamma$, and Eq.\ (\ref{eq:fact2})
reduces to $\mu_r \sim \Gamma^r$.  Substitution into Eq.\ 
(\ref{eq:main}) yields the power--law behavior
\begin{eqnarray}
{\cal P}(\mu_r) \sim \mu_r^{{(\beta M) / (2r)}-1} ,
\end{eqnarray}
for $\mu_r \to 0$.  A special case of this result is $M=\beta=1$ for
which the distribution of the first and second moment diverge as
$\mu_1^{-1/2}$ and $\mu_2^{-3/4}$ for $\mu_1$, $\mu_2 \to 0$,
respectively.

The third case is the short--time regime $t \ll t_c$ where $t_c$ is
the correlation time of the intensity fluctuations of the cavity
mode.  During counting intervals that short the field intensity
cannot vary appreciably. Therefore, the integrated intensity $I$
becomes the product of $t$ with the instantaneous photon number
$\hat{n}=b^\dagger b$ in the cavity. The factorial moments of the
output take the simple form
\begin{eqnarray}
\mu_r = (\Gamma t)^r \langle: \hat{n}^r \! :\rangle \, ,
\label{eq:fact3}
\end{eqnarray}
and one easily verifies that the normally ordered moment $\langle:
\hat{n}^r \! :\rangle$ occuring here reduces to the $r$th factorial
moment of the stationary photon number distribution of the lasing
mode. The latter has the form \cite{Wal94,Man95}
\begin{eqnarray}
\label{eq:distin}
P_n = {\cal N} \ { [{\cal A} n_s / {\cal C}]^{n+n_s} \over
\! \! (n+n_s)! }\;.
\end{eqnarray}
Note that $P_n$ depends on the three laser parameters ${\cal A}$,
${\cal B}$, and ${\cal C}$; in particular, the nonlinearity ${\cal
  B}$ enters through the so--called saturation photon number $n_s =
{\cal A} / {\cal B}$. The symbol ${\cal N}$ represents a
normalization constant.

\begin{figure}[t]
\begin{center}
\leavevmode
\vspace*{-1.1cm}
\epsfxsize = 7.3cm
\epsfbox{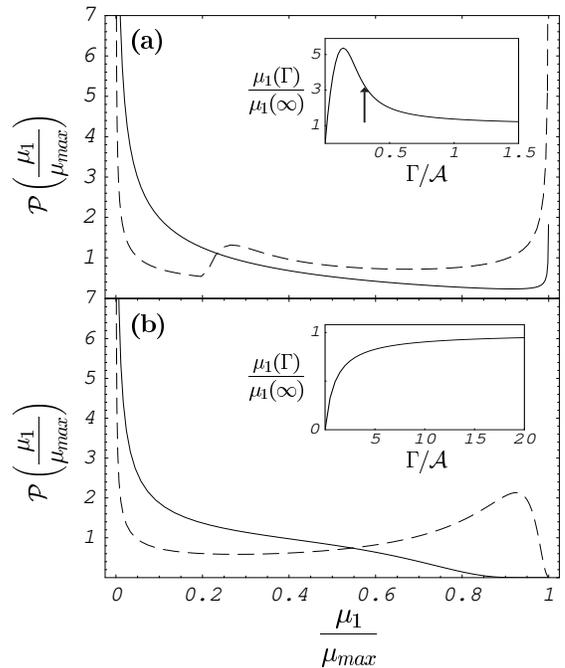}
\end{center}
\vspace*{0.9cm}
\caption{Distribution ${\cal P}(\mu_1 / \mu_{\rm  max})$ as a
  function of the dimensionless mean photocount $\mu_1 / \mu_{\rm
    max}$ for $\beta M=1$ and four sets of laser parameters. Rates
  are given in units of ${\cal A} \equiv 1$, the nonlinearity is
  ${\cal B}= 0.005$. (a) $\kappa= 0.7$, $\overline{\Gamma} = 0.02$
  (solid line); $\kappa= 0.7$, $\overline{\Gamma} = 0.2$ (dashed
  line). (b) $\kappa= 2.0$, $\overline{\Gamma} = 0.5$ (solid line);
  $\kappa= 2.0$, $\overline{\Gamma} = 4.0$ (dashed line). The insets
  in (a) and (b) show $\mu_1$ as a function of $\Gamma /{\cal A}$.
  The arrow indicates the threshold--value of $\Gamma$ below which
  lasing takes place.}
\end{figure}

The calculation of the mean photocount $\mu_1$ reduces to a
steady--state average for arbitrary counting time. We calculated
$\mu_1$ numerically as a function of $\Gamma$ using the distribution
(\ref{eq:distin}).  Substituting the result $\mu_1(\Gamma)$ into
Eq.\ (\ref{eq:main}) and carrying out the integration over $\Gamma$,
we obtained ${\cal P}(\mu_1)$.  The results for time--reversal
invariant cavities connected to a single--mode waveguide are plotted
in Fig.~2, for $\beta M=1$ and four different sets of parameters.
The two distributions in Fig.\ 2(a) correspond to ${\cal A}>
\overline{\cal C}$, i.e.\ to lasers above threshold in the ensemble
average. By contrast, ${\cal A}< \kappa$ for the distributions of
Fig.\ 2(b); all lasers of those ensembles are below threshold
irrespective of the escape rate $\Gamma$. We note that all
distributions are strongly non--Gaussian. They are all peaked as
$\mu_1^{-1/2}$ at small photocount, in accord with the general
argument presented above for the asymptotics at $\mu_1 \to 0$. Two
further features spring to the eye and demand explanation: First,
above threshold but not below we encounter an additional peak at
maximum photocount; second, for certain cases (dashed lines in Fig.\ 
2) the distribution ${\cal P}(\mu_1)$ displays a shoulder for
sub--maximal $\mu_1$.

The origin of these structures lies in the $\Gamma$--dependence of
the mean photocount $\mu_1$. That dependence is depicted in the
insets and seen to be qualitatively different above and below
threshold.  While $\mu_1$ increases monotonically with $\Gamma$ in
the below--threshold case, it develops a maximum at an intermediate
value of $\Gamma$ when the laser is above threshold.  This behavior
can be understood from the simple analytic expressions
\begin{eqnarray}\label{eq:cases}
\mu_1/t=\cases{
 \Gamma n_s\frac{{\cal A}-{\cal C}}{{\cal C}}  & for
${\cal C}=\Gamma+\kappa<{\cal A}\;,$ \cr
 \frac{\Gamma {\cal A}}{{\cal C}-{\cal A}}& for
${\cal C}=\Gamma+\kappa>{\cal A}\;.$ \cr}
\end{eqnarray}
that follow from the photon number distribution (\ref{eq:distin})
sufficiently far from threshold. According to Eq.\ (\ref{eq:cases}),
$\mu_1$ rises linearly with $\Gamma$ out of the origin and
approaches an asymptotic plateau for very large $\Gamma$, as visible
in Fig.\ 2.  The rise to the plateau is monotonic when ${\cal
  A}<\kappa$ since the below-threshold case of (\ref{eq:cases})
applies for all values of $\Gamma$; the maximum photocount $\mu_{\rm
  max}$ is then just the plateau value. The laser of Fig.\ 2(a) is
above threshold for small $\Gamma$ but below for large $\Gamma$. The
maximum photocount $\mu_{\rm max}$ then arises for an intermediate
value $\Gamma^*$. A simple argument can be employed to determine the
border between the below-- and the above--threshold ensemble. The
argument follows from the observation that $\mu_1(\Gamma)$
approaches its plateau--value from above for the above--threshold
ensemble while the plateau is approached from below when the
ensemble is below threshold.  From the lower case of
Eq.~(\ref{eq:cases}), which becomes exact for $\Gamma\to\infty$,
this yields the threshold condition ${\cal A} = \kappa$. To estimate
the value of $\Gamma^*$ we may employ the above--threshold case of
Eq.~(\ref{eq:cases}) and find $\Gamma^*=\sqrt{{\cal
    A}\kappa}-\kappa$.

Based on this understanding of $\mu_1(\Gamma)$ one can appreciate
the above--threshold peak of ${\cal P}(\mu_1)$ at $\mu_1=\mu_{\rm
  max}$. Substituting $\mu_1(\Gamma)= \mu_{\rm max} +
\frac{1}{2}\mu''(\Gamma^*)(\Gamma-\Gamma^*)^2$ in the photocount
distribution (\ref{eq:main}) we find that ${\cal P} (\mu_1)$ has a
square--root singularity $|\mu_1 - \mu_{\rm max}|^{-1/2}$ which is
precisely the peak depicted in Fig.~2(a).  Clearly, no such peak can
arise in the below--threshold case of Fig.~2(b) as the photocount
increases monotonically with $\Gamma$.  Inspecting Eqs.\ 
(\ref{eq:distgam}), (\ref{eq:main}) in the limit $\Gamma \to \infty$
we rather find that ${\cal P}(\mu_1)$ vanishes when $\mu_1$
approaches $\mu_{\rm max}$.

The shoulders at sub--maximal photocount are caused by amplified
spontaneous emission below the laser threshold.  Formally, the
shoulders arise from the asymptotic plateau of $\mu_1$ for large
$\Gamma$.  The definition (\ref{eq:main}) immediately implies
enhanced probability for photocounts $\mu_1$ in the vicinity of that
plateau.  Note that the shoulder is invisible for the distribution
shown as a solid line in Fig.~2(a), since for this distribution
large values of $\Gamma$ are strongly suppressed by the small mean
value $\overline{\Gamma}$.  Further note that the shoulders for both
curves in Fig.\ 2(b) lie closer to $\mu_{\rm max}$ than for the
dashed curve in Fig.\ 2(a) since $\mu \to \mu_{\rm max}$ coincides
with $\Gamma \to \infty$ in the regime below threshold.

In contrast to the mean photocount which can be expressed in terms
of the stationary distribution of the laser, all $\mu_r$ with $r \ge
2$ involve dynamical information through correlation functions with
$r-1$ time arguments. We defer the discussion of the higher
factorial moments and their distribution to a separate publication
\cite{fut}.

We finally compare our results with related fluctuation phenomena in
other areas of physics and discuss possible experimental tests. In
nuclear physics the Porther--Thomas distribution describes
level--width fluctuations in neutron scattering \cite{Guh98}. The
amplitude fluctuations of Coulomb blockade oscillations in
semiconductor quantum dots are also of the Porther--Thomas type
\cite{Jal92}. In both cases, as well as for the chaotic lasers
studied in this paper, the fluctuations result from the chaotic
nature of wave functions.  However, in chaotic lasers new
interesting features arise due to the interplay of wave chaos with
the nonlinear dynamics of the laser. As a consequence of this
interplay, the distribution of the mean photocount can strongly
deviate from the Porter--Thomas distribution.  To test the predicted
mode--to--mode fluctuations experimentally, one must study the
photocount statistics for an ensemble of chaotic modes. It seems
feasible to generate such ensembles in tunable lasers e.g.\ by shape
variations, or by the injection and displacement of artificial
scatterers in the case of microwave cavities.

Support by the Sonderforschungsbereich ``Unordnung und gro\ss e
Fluktuationen'' der Deutschen Forschungsgemeinschaft is gratefully
acknowledged.


\vspace*{-0.0cm}

\begin{references}

\vspace*{-1.7cm}

\bibitem{Wal94} D.\ F.\ Walls and G.\ J.\ Milburn, {\it Quantum
    Optics} (Springer, Berlin, 1994).

\bibitem{Man95} L.\ Mandel and E.\ Wolf, {\it Optical Coherence and
    Quantum Optics} (Cambridge University Press, Cambridge, 1995).

\bibitem{Law94} M.\ N.\ Lawandy, R.\ M.\ Balachanfran, A.\ S.\ L.\
Gomes, and E.\ Sauvain, Nature (London) {\bf 368}, 436 (1994).

\bibitem{Cao99} H.\ Cao, Y.\ G.\ Zhao, S.\ T.\ Ho, E.\ W.\ Seelig,
Q.\ H.\ Wang, and R.\ P.\ H.\ Chang, Phys.\ Rev.\ Lett.\ {\bf 82},
2278 (1999).

\bibitem{Gma98} C.\ Gmachl, F.\ Capasso, E.\ E.\ Narimanov, J.\ U.\ 
  N\"ockel, A.\ D.\ Stone, J.\ Faist, D.\ L.\ Sivco, and A.\ Y.\ 
  Cho, Science {\bf 280}, 1556 (1998).

\bibitem{She90} {\it Scattering and Localization of Classical Waves
    in Random Media}, edited by P.\ Sheng (World Scientific,
  Singapore, 1990).

\bibitem{Gru96} T.\ Gruner and D.--G.\ Welsch, Phys.\ Rev.\ A {\bf
54}, 1661 (1996).

\bibitem{Art97} M.\ Artoni and R.\ Loudon, Phys.\ Rev.\ A {\bf 55},
1347 (1997).

\bibitem{Bee98} C.\ W.\ J.\ Beenakker, Phys.\ Rev.\ Lett.\ {\bf 81},
  1829 (1998).

\bibitem{Mis00} E.\ G.\ Mishchenko, M.\ Patra, and C.\ W.\ J.\
Beenakker, cond-mat/0003262.

\bibitem{Bee00} C.\ W.\ J.\ Beenakker, M.\ Patra, and P.\ W.\
 Brouwer, Phys.\ Rev.\ A {\bf 61}, 051801(R) (2000).

\bibitem{Berry} M.\ V.\ Berry, J.\ Phys.\ A {\bf 10}, 2083 (1977).

\bibitem{Bibel} F.\ Haake {\it Quantum Signatures of Chaos}, 2nd
  edition (Springer, Berlin, 2000).

\bibitem{Gar85} C.\ W.\ Gardiner and M.\ J.\ Collet, Phys.\ Rev.\ A
  {\bf 31}, 3761 (1985).

\bibitem{Guh98} T.\ Guhr, A.\ M\"uller--Groehling, and H.\ A.\
  Weidenm\"uller, Phys.\ Rep.\ {\bf 299}, 189 (1998).

\bibitem{beta} The case $\beta=2$ can be realized in microwave
  cavities containing magnetized ferrites.

\bibitem{fut} G.\ Hackenbroich, C.\ Viviescas, B.\ Elattari, and F.\
  Haake, unpublished.

\bibitem{Jal92} R.\ A.\ Jalabert, A.\ D.\ Stone, and Y.\ Alhassid,
  Phys.\ Rev.\ Lett.\ {\bf 68}, 3468 (1992).


\end{references}
\end{document}